# Towards Simple and Useful One-Time Programs in the Quantum Random Oracle Model

**Lev Stambler**
levstamb@umd.edu
University of Maryland, College Park,
NeverLocal Ltd.

January 12, 2026

**Abstract**

We construct simulation-secure one-time memories (OTM)[1] in the random oracle model, and present a plausible argument for their security against quantum adversaries with bounded and adaptive depth[2]. Our contributions include:

- **A Simple scheme**: We use only single-qubit Wiesner states and conjunction obfuscation (constructible from LPN): no complex entanglement or quantum cryptography is required.
- **New POVM bound**: We prove that any measurement achieving $(1-\epsilon)$ success on one basis has conjugate-basis guessing probability at most $\frac{1}{2^m} + O\left(\epsilon^{\frac{1}{4}}\right)$.
- **Adaptive depth security**: Via an informal application of a lifting theorem, we conjecture security against adversaries with polynomial quantum circuit depth between random oracle queries.
- **Practical path**: Our construction requires only single-qubit operations and classically-instantiable obfuscation, pointing toward near-term implementation.

Security against adaptive quantum adversaries captures many realistic attacks on OTMs built from single-qubit states; our work thus paves the way for practical and truly secure one-time programs. Moreover, depth bounded adaptive adversarial models may allow for encoding one-time memories into error corrected memory states, opening the door to implementations of one-time programs which persist for long periods of time.

## 1 Introduction

One-time programs, first introduced by Goldwasser, Kalai, and Rothblum [10], are a powerful cryptographic primitive that allows a program to be run only once. In the quantum setting, one-time programs have been studied by Broadbent, Gutoski, and Stebila [6]. In both the quantum and classical setting, one-time programs cannot be securely constructed from standard cryptographic assumptions alone [6, 10]. Thus further assumptions are required to realize one-time programs.

[1] Which in turn imply one-time programs [10]

[2] Here adaptive indicates that the quantum adversary can interleave classical computation into its quantum computation, depending on measurement outputs. Formally, this is the complexity class $\mathsf{BPP}^{\mathsf{QNC}^{\mathsf{BPP}}}$ as per Arora et al. [1].

In this work, we focus on the security of one-time programs against adversaries with limited, though adaptive, quantum depth: specifically, adversaries in the complexity class $\mathsf{BPP}^{\mathsf{QNC}_d^{\mathsf{BPP}}}$ as defined by Arora et al. [1]. Such adversaries can interleave polynomial-time classical computation with quantum circuits of polynomial depth, allowing them to maintain quantum memory across oracle queries while restricting the overall quantum processing power. This is a remarkably *weak* assumption on the adversary: the only restriction is bounded quantum circuit depth per round, while the adversary retains unbounded classical computation, arbitrarily many adaptive rounds, and quantum coherence throughout each bounded-depth circuit. Moreover, bounded depth is physically natural for input states that are *not* fault-tolerantly encoded — without error correction, noise accumulates with circuit depth, so an adversary's coherent processing power over such states is inherently limited. We believe this setting is realistic not only for current hardware, but will remain so in the future. Many current one-time program constructions require the adversary to either measure their quantum state within a bounded time frame (e.g. [11, 17]) or place some other restriction on quantum memory [9, 13–15] which limit the usefulness of the OTP. On the other hand, our scheme allows the receiver to *hold* a quantum one-time program for later use while being confident that an adversary with limited quantum processing power cannot break its security[3].

## 1.1 Our Contributions

We construct one-time memories from tensor products of independent single-qubit Wiesner states and conjunction obfuscation (instantiable from LPN [3][4]), avoiding multi-qubit entanglement, hidden subspace states, or indistinguishability obfuscation required by prior work [13, 17].

The security of our scheme rests on a new POVM bound for $m$-qubit conjugate coding (Section 3): any measurement identifying computational basis states with probability $1 - \epsilon$ allows guessing the conjugate string with probability at most $\frac{1}{2^m} + O\left(\epsilon^{\frac{1}{4}}\right)$. This bound is *sequential*—it holds even when the adversary learns the basis choice after measuring—and generalizes prior single-qubit results [19, 20].

Then, using this bound, we first construct an OTM and prove its security against an adversary which can only make classical queries to a random oracle in Section 4. We then argue, via an informal application of the lifting framework of Arora et al. [1], that our OTM should extend to adaptive quantum adversaries with polynomial quantum depth between oracle queries (Section 5).

**Technical Overview**

The security of our scheme rests on a new information-theoretic bound for conjugate coding (Section 3). We show that any POVM achieving success probability $(1 - \epsilon)$ on identifying $m$ qubits in the computational basis allows guessing the string encoded in the conjugate (Hadamard) basis with probability at most $1/2^m + O\left(\epsilon^{1/4}\right)$. Crucially, this bound is *sequential*:

[3] To see potential use-cases for "short-lived" one-time programs, see the author's prior work [18].

[4] LPN, Learning Parity with Noise, is a well-studied computational assumption believed to be quantum-resistant. The key idea is that given random linear equations with some noise, it is hard to recover the underlying secret vector. Ref. [3] shows how to construct conjunction obfuscation from $O$(security parameter) LPN cipher-text, rendering the construction quite practical.

it holds even when the adversary learns which basis to guess *after* performing their measurement. This generalizes prior single-qubit results [19, 20] to the multi-qubit setting, and for $m = \Theta(\lambda)$ qubits, makes conjugate-basis guessing exponentially hard.

Our OTM construction (Section 4) uses this bound as follows. The sender prepares $n$ independent $m$-qubit conjugate coding states, each encoding a uniformly random secret $s_i \in \{0,1\}^m$. A random subset $\Theta_X$ of positions uses the $X$ basis and the complementary set $\Theta_Z$ uses the $Z$ basis. Two conjunction obfuscations, each checking whether the hashed measurement outcomes for its positions match the expected values, hide which positions belong to which basis and output a key $k_\alpha$ upon successful evaluation. The messages are encrypted as $c_\alpha = k_\alpha \oplus m_\alpha$, so recovering $m_\alpha$ requires evaluating the conjunction $\mathcal{O}_\alpha$. The random oracle forces the adversary to commit to specific measurement outcomes via classical queries. And, our POVM bound guarantees that measuring correctly in one basis leaves conjugate-basis secrets with high min-entropy. Thus, the adversary cannot produce valid oracle queries for both conjunctions, enforcing one-time execution.

Finally, we argue that classical-query security should lift to the adaptive quantum depth setting (Section 5) via the lifting framework of Arora et al. [1]. Their framework shows that $\mathsf{BPP}^{\mathsf{QNC}_d^{\mathsf{BPP}}}$ adversaries, which make oracle queries to a specially constructed random oracle, cannot coherently make quantum queries to these oracles. So, such adversaries' oracle queries effectively collapse to classical transcripts, which suggests our classical-query security proof should carry through. We conjecture that this yields an OTM that honest users can store in quantum memory for later use, while remaining secure against realistic adversaries with limited quantum processing power. Importantly, the random oracle can be concretely instantiated in a practical manner: informally, the special random oracle $\mathcal{H}$ required by Arora et al. can be instantiated by composing independent hash functions, $\mathcal{H}(x) = H_{O(\lambda)}\big(H_{O(\lambda)-1}(...H_1(x)...)\big)$, where each $H_i$ is modeled as an independent random oracle (see [1] for details).

## 1.2 Related Work

One-time memories (OTMs) and one-time programs (OTPs) have a rich history spanning both classical and quantum settings. Though impossible in the standard model, OTMs/OTPs can be constructed under various additional assumptions. Many lines of work use hardware assumptions such as stateless hardware tokens [5, 7, 19][5], physical assumptions such as isolated qubits [13–15], restrict adversarial capabilities such as bounded/noisy quantum storage [2, 4, 9], or bounded quantum depth in one capacity or another [11, 17, 20] to construct OTMs/OTPs.

Though all of these notions are interesting and have made substantial progress, we believe that they have limitations in terms of practicality and applicability. Hardware tokens are difficult to implement in practice and have had security vulnerabilities [8, 12, 16]. The isolated qubits model, bounded storage, and depth bounded assumptions which require limited coherence time are physically motivated, but restrict the capabilities of the one-time programs. Mainly, we would hope to be able to store our OTPs in quantum memory so that they can be used at any time in the future.

[5]Hardware tokens can alternatively be thought of as stateless secure hardware such as Trusted Execution Environments (TEEs)

The work of Sekii and Nishide [17] is most related to ours, but they use a "measure-and-prepare" adversarial model (no quantum memory between rounds) and require both hidden subspace states and iO.

In our work, we allow depth-bounded quantum adversaries that can maintain quantum memory across oracle queries, as long as the inter-query quantum circuit depth is polynomially bounded. We believe that this is a natural model for auxiliary quantum information processing: if a quantum state is received from a third party in a non-fault tolerant manner, it is likely that the quantum processing capabilities are limited.

### 1.3 Roadmap

The rest of the paper is organized as follows. Section 2 recalls the necessary quantum and cryptographic definitions. In Section 3, we prove our main technical lemma bounding the guessing probability for conjugate-basis measurements. In Section 4, we present our one-time memory scheme and prove classical-query security. In Section 5, we discuss how to informally lift this to polynomial-depth quantum security via the Arora et al. framework. Finally, we conclude with future directions.

## 2 Preliminaries

We recall the necessary quantum and cryptographic definitions. Mainly, we will use conjugate coding, conjunction obfuscation, and one-time memory. Further, we define our adversarial model as the complexity class $\mathsf{BPP}^{\mathsf{QNC}^{\mathsf{BPP}}}$ following Ref. [1].

### 2.1 Quantum Preliminaries

**Definition 2.1** *(POVM)*: A *positive operator-valued measure* (POVM) on a $d$-dimensional Hilbert space $\mathcal{H}$ is a collection of positive semidefinite operators $\{M_i\}_{i\in\mathcal{I}}$ satisfying $\sum_{i\in\mathcal{I}} M_i = \mathbb{I}$. Given a quantum state $\rho$, the probability of outcome $i$ is $\mathrm{Tr}(M_i\rho)$.

**Definition 2.2** *(Conjugate Coding (Wiesner States))*: For $m \in \mathbb{N}$, we define the two conjugate bases on $(\mathbb{C}^2)^{\otimes m}$:

- **Computational basis ($Z$):** The states $|z\rangle$ for $z \in \{0,1\}^m$
- **Hadamard basis ($X$):** The states $|\psi_x\rangle = H^{\otimes m}|x\rangle$ for $x \in \{0,1\}^m$

A *Wiesner state* with secret $s \in \{0,1\}^m$ and basis $\theta \in \{X, Z\}$ is:

$$|s\rangle_\theta = \begin{cases} |s\rangle & \text{if } \theta = Z \\ H^{\otimes m}|s\rangle & \text{if } \theta = X \end{cases} \tag{1}$$

### 2.2 Conjunction Obfuscation

**Definition 2.3** *(Distributional VBB Obfuscation with Auxiliary Input)*: Let $\mathcal{C} = \{\mathcal{C}_\lambda\}_{\lambda\in\mathbb{N}}$ be a family of circuit classes and let $\mathcal{D} = \{\mathcal{D}_\lambda\}_{\lambda\in\mathbb{N}}$ be a family of distributions over $\mathcal{C}$. A PPT algorithm ObfConj is a *distributional VBB obfuscator with auxiliary input* for $(\mathcal{C}, \mathcal{D})$ if the following properties hold:

1. **Functionality**: For all $\lambda \in \mathbb{N}$ and all $C \in \mathcal{C}_\lambda$, the circuit $\mathrm{ObfConj}(1^\lambda, C)$ computes the same function as $C$.
2. **Polynomial slowdown**: There exists a polynomial $p$ such that for all $C \in \mathcal{C}_\lambda$, $|\mathrm{ObfConj}(1^\lambda, C)| \le p(|C|)$.

3. **Distributional VBB security with auxiliary input**: For any family of auxiliary inputs $\{\text{aux}_\lambda\}_{\lambda\in\mathbb{N}}$ where $|\text{aux}_\lambda| \le \text{poly}(\lambda)$, and any PPT adversary $\mathcal{A}$, there exists a PPT simulator $\mathcal{S}$ such that for $C \leftarrow \mathcal{D}_\lambda$:
$$\left|\Pr[\mathcal{A}(1^\lambda, \text{ObfConj}(1^\lambda, C), \text{aux}_\lambda) = 1] - \Pr\left[\mathcal{S}^C(1^\lambda, \text{aux}_\lambda) = 1\right]\right| \le \text{negl}(\lambda). \quad (2)$$

For our construction, we instantiate ObfConj with the conjunction obfuscator of Bartusek et al. [3], which achieves distributional VBB security with classical auxiliary input for the class of conjunctions under the Learning Parity with Noise (LPN) assumption.

We will need to allow for quantum auxiliary input in our security proof. Given the use of post-quantum LPN to construct conjunction obfuscation, we expect that the obfuscator should remain secure even with quantum auxiliary input as long as the accepting input is high entropy given the auxiliary input. We thus extend the security of the above obfuscator with the following conjecture.

**Conjecture 2.1** *(Distributional VBB with Quantum Auxiliary Input)*: The conjunction obfuscator of [3] satisfies distributional VBB security even when the auxiliary input $\text{aux}_\lambda$ is a polynomial-size quantum state, and the adversary and simulator are QPT algorithms.

## 2.3 One-Time Memory

A one-time memory (OTM) is a cryptographic primitive that allows a sender to encode two messages such that a receiver can retrieve exactly one of them [10]. Informally, a one-time memory token for messages $(m_0, m_1)$ allows the holder to learn exactly one of the two messages, with no information about the other. We use a definition of simulation security following Liu [11].

**Definition 2.4** *(Simulation Security)*: A memory token scheme (Token.Gen, Token.Eval) in the QROM is *simulation-secure* against $d$-depth quantum adversaries if for every message length $\ell$, every $d$-depth quantum adversary $\mathcal{A}$, and every inverse polynomial $\gamma(\cdot)$, there exists an efficient quantum simulator Sim making at most one classical query to $g^{m_0,m_1} : b \mapsto m_b$ such that:
$$\mathcal{A}\left(\text{Token.Gen}^f(1^\lambda, m_0, m_1), \texttt{aux}\right) \underset{\gamma(\lambda)}{\approx} \text{Sim}^{g^{m_0,m_1}}\left(1^\lambda, \texttt{aux}\right) \quad (3)$$
where $f$ is a uniformly random function (the random oracle).

## 2.4 Adversarial Model: $\mathsf{BPP}^{\mathsf{QNC}^{\mathsf{BPP}}}$

We model our adversaries as belonging to the complexity class $\mathsf{BPP}^{\mathsf{QNC}^{\mathsf{BPP}}}$ following Ref. [1]. Intuitively, this class captures polynomial-time classical computation with access to quantum computation of bounded depth.

**Definition 2.5** *($\mathsf{QNC}_d^{\mathsf{BPP}}$)*: A computation is in $\mathsf{QNC}_d^{\mathsf{BPP}}$ if it consists of $d$ quantum layers $U_1, ..., U_d$ interleaved with polynomial-time classical computations $B_1, ..., B_{d-1}$:
$$U_1 \to B_1 \to U_2 \to B_2 \to \cdots \to B_{d-1} \to U_d \quad (4)$$
Each $U_i$ is a layer of polynomially many parallel one- and two-qubit gates. Crucially, **quantum state persists** across the classical computations: each $B_i$ receives classical input (e.g., from intermediate measurements or the original input) and produces classical output that may influence subsequent layers, but the unmeasured quantum register passes coherently from $U_i$ to $U_{i+1}$.

**Definition 2.6** ($\mathsf{BPP}^{\mathsf{QNC}_d^{\mathsf{BPP}}}$): A language $L$ is in $\mathsf{BPP}^{\mathsf{QNC}_d^{\mathsf{BPP}}}$ if there exists a polynomial $m(\cdot)$ and a family of computations consisting of $m$ adaptive rounds:

$$A_1 \to Q_1 \to A_2 \to Q_2 \to \cdots \to A_m \to Q_m \to A_{m+1} \tag{5}$$

where each $A_i$ is a polynomial-time classical algorithm and each $Q_i$ is a $\mathsf{QNC}_d^{\mathsf{BPP}}$ circuit as defined above. Between rounds, all quantum state is measured: the classical algorithm $A_{i+1}$ receives the measurement outcomes from $Q_i$ along with all prior classical information, and prepares the classical input for $Q_{i+1}$. The final algorithm $A_{m+1}$ outputs the decision bit.

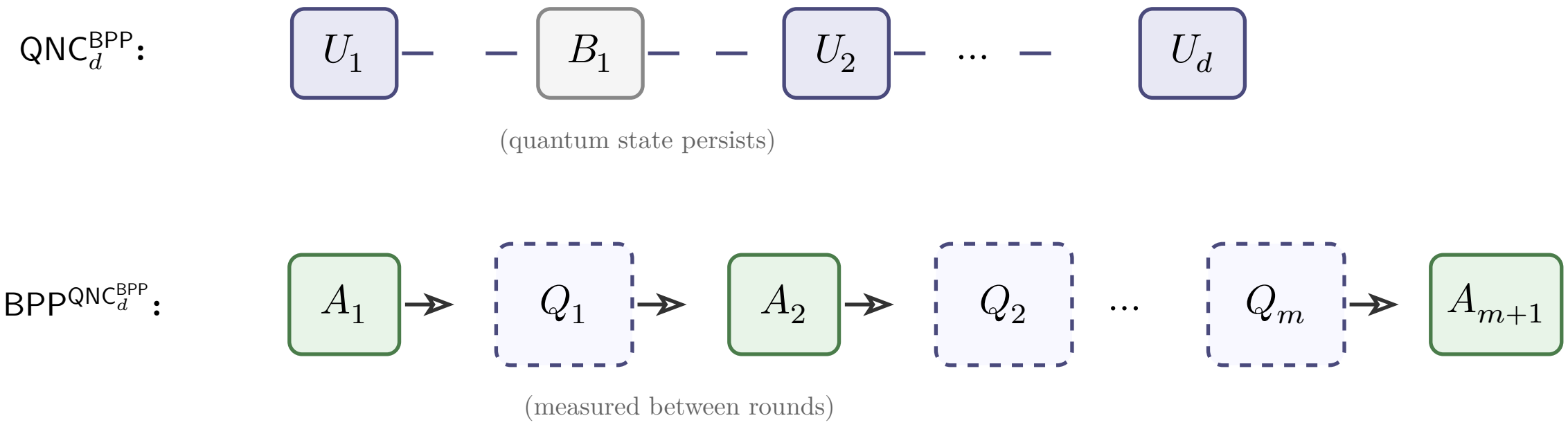

Figure 1: Top: $\mathsf{QNC}_d^{\mathsf{BPP}}$ circuit with $d$ quantum layers ($U_i$) and classical subroutines ($B_i$) where quantum state passes coherently. Bottom: $\mathsf{BPP}^{\mathsf{QNC}_d^{\mathsf{BPP}}}$ with $m$ rounds of classical computation ($A_i$) invoking $\mathsf{QNC}_d^{\mathsf{BPP}}$ circuits ($Q_i$), with measurements between rounds.

# 3 POVM Bounds

We now establish that any measurement strategy achieving high success on one basis yields low guessing probability in the conjugate basis.

Importantly, our bound is "sequential" in nature: it allows for an adversary to first measure and then *learn* which basis the original state was prepared in before guessing. This is similar to the author's prior work, but generalized to arbitrary conjugate coding states of dimension $2^m$ rather than single-qubit states [19, 20].

**Theorem 3.1** (Sequential Conjugate Coding Security): Let $\{M_z\}_{z\in\{0,1\}^m}$ be a POVM on an $m$-qubit system. Suppose the POVM correctly identifies computational basis states with high probability:

$$\mathop{\mathbb{E}}_{z\leftarrow\{0,1\}^m}[\mathrm{Tr}(M_z|z\rangle\langle z|)] \geq 1-\epsilon. \tag{6}$$

Then for a uniformly random Hadamard basis state $|\psi_x\rangle = H^{\otimes m}|x\rangle$, the optimal guessing probability for $x$ given the POVM outcome is bounded:

$$\max_f \mathop{\mathbb{E}}_{x\leftarrow\{0,1\}^m}\left[\mathrm{Tr}\left(M_{f(x)}|\psi_x\rangle\langle\psi_x|\right)\right] \leq \frac{1}{2^m} + O\left(\epsilon^{\frac{1}{4}}\right) \tag{7}$$

where the maximum is over all functions $f : \{0,1\}^m \to \{0,1\}^m$ attempting to guess $x$ from the measurement outcome.

Before proving the theorem, we set up some notation and intermediate results. Let $D = 2^m$ denote the dimension and define:

$$\mathcal{G} = \{i \in [D] : \mathrm{Tr}(M_i|i\rangle\langle i|) \geq 1-\sqrt{\epsilon}, M_i \in \mathcal{M}_z\} \tag{8}$$

as the set of "good" measurement elements, and $\mathcal{B} = [D] \setminus \mathcal{G}$ as the "bad" elements. By Markov's inequality applied to the assumption $\mathbb{E}_z[\mathrm{Tr}(M_z|z\rangle\langle z|)] \geq 1-\epsilon$, we have

$$\frac{|\mathcal{B}|}{D} \leq \sqrt{\epsilon}. \tag{9}$$

**Lemma 3.1** *(Good Element Approximation)*: For $M_i \in \mathcal{G}$, the POVM element is close to the corresponding projector in trace norm:

$$\|M_i - |i\rangle\langle i|\|_1 \leq 2\epsilon^{\frac{1}{4}}. \tag{10}$$

*Proof*: For $M_i \in \mathcal{G}$, we have $\mathrm{Tr}(M_i|i\rangle\langle i|) = M_i[i,i] \geq 1-\sqrt{\epsilon}$. Let $\Pi_i = |i\rangle\langle i|$. Using the Fuchs–van de Graaf inequality relating trace distance and fidelity:

$$\|M_i - \Pi_i\|_1 \leq 2\sqrt{1 - F(M_i, \Pi_i)^2} \tag{11}$$

where the fidelity is $F(M_i, \Pi_i) = \mathrm{Tr}\left[\sqrt{\sqrt{\Pi_i} M_i \sqrt{\Pi_i}}\right]$. Since $\Pi_i$ is rank-1:

$$\begin{aligned} F(M_i, \Pi_i) &= \mathrm{Tr}\left[\sqrt{|i\rangle\langle i|M_i|i\rangle\langle i|}\right] \\ &= \mathrm{Tr}\left[\sqrt{M_i[i,i] \cdot |i\rangle\langle i|}\right] = \sqrt{M_i[i,i]} \geq \sqrt{1-\sqrt{\epsilon}}. \end{aligned} \tag{12}$$

Thus $1 - F^2 \leq 1 - (1-\sqrt{\epsilon}) = \sqrt{\epsilon}$, giving:

$$\|M_i - \Pi_i\|_1 \leq 2\sqrt{\sqrt{\epsilon}} = 2\epsilon^{\frac{1}{4}}. \tag{13}$$

■

**Lemma 3.2** *(Measurement Probability Bound)*: For any $x \in \{0,1\}^m$ and $M_i \in \mathcal{G}$:

$$\mathrm{Tr}(M_i|\psi_x\rangle\langle\psi_x|) \leq \frac{1}{D} + 2\epsilon^{\frac{1}{4}}. \tag{14}$$

*Proof*: By Hölder's inequality for trace norm:

$$\begin{aligned} \|\mathrm{Tr}(M_i|\psi_x\rangle\langle\psi_x|) - \mathrm{Tr}(|i\rangle\langle i||\psi_x\rangle\langle\psi_x|)\|_1 &= \|\mathrm{Tr}((M_i - |i\rangle\langle i|)|\psi_x\rangle\langle\psi_x|)\|_1 \\ &\leq \|M_i - |i\rangle\langle i|\|_1 \cdot \||\psi_x\rangle\langle\psi_x|\|_\infty. \end{aligned} \tag{15}$$

Since $\||\psi_x\rangle\langle\psi_x|\|_\infty = 1$ and by Lemma 3.1:

$$\begin{aligned} \mathrm{Tr}(M_i|\psi_x\rangle\langle\psi_x|) &= \mathrm{Tr}[(M_i - |i\rangle\langle i| + |i\rangle\langle i|) \quad |\psi_x\rangle\langle\psi_x|] \\ &\leq \mathrm{Tr}(|i\rangle\langle i||\psi_x\rangle\langle\psi_x|) + \|M_i - |i\rangle\langle i|\|_1 \\ &\leq \mathrm{Tr}(|i\rangle\langle i||\psi_x\rangle\langle\psi_x|) + 2\epsilon^{\frac{1}{4}} = |\langle i|\psi_x\rangle|^2 + 2\epsilon^{\frac{1}{4}}. \end{aligned} \tag{16}$$

Since $|\psi_x\rangle = H^{\otimes m}|x\rangle$ is the uniform superposition with phase $(-1)^{i\cdot x}$:

$$|\langle i|\psi_x\rangle|^2 = |\frac{1}{\sqrt{D}}(-1)^{i\cdot x}|^2 = \frac{1}{D}. \tag{17}$$

Therefore $\mathrm{Tr}(M_i|\psi_x\rangle\langle\psi_x|) \leq \frac{1}{D} + 2\epsilon^{\frac{1}{4}}$. ■

We now prove the main theorem:

*Proof*: Let $f : \{0,1\}^m \to \{0,1\}^m$ be any guessing function. The guessing probability is:

$$\mathbb{E}_x\left[\mathrm{Tr}\left(M_{f(x)}|\psi_x\rangle\langle\psi_x|\right)\right] = \frac{1}{D}\sum_x \mathrm{Tr}\left(M_{f(x)}|\psi_x\rangle\langle\psi_x|\right). \tag{18}$$

We partition based on whether $f(x) \in \mathcal{G}$ or $f(x) \in \mathcal{B}$:

$$\frac{1}{D}\sum_x \mathrm{Tr}\left(M_{f(x)}|\psi_x\rangle\langle\psi_x|\right) = \frac{1}{D}\sum_{x:f(x)\in\mathcal{G}} \mathrm{Tr}\left(M_{f(x)}|\psi_x\rangle\langle\psi_x|\right) + \frac{1}{D}\sum_{x:f(x)\in\mathcal{B}} \mathrm{Tr}\left(M_{f(x)}|\psi_x\rangle\langle\psi_x|\right). \tag{19}$$

**Good elements:** By Lemma 3.2, for each $x$ with $f(x) \in \mathcal{G}$:

$$\mathrm{Tr}\Big(M_{f(x)}|\psi_x\rangle\langle\psi_x|\Big) \le \frac{1}{D} + 2\epsilon^{\frac{1}{4}}. \tag{20}$$

Summing over all such $x$ (at most $D$ terms):

$$\frac{1}{D}\sum_{x:f(x)\in\mathcal{G}} \mathrm{Tr}\Big(M_{f(x)}|\psi_x\rangle\langle\psi_x|\Big) \le \frac{1}{D} + 2\epsilon^{\frac{1}{4}}. \tag{21}$$

**Bad elements:** We bound the total probability mass on bad elements. Using $\mathbb{E}_x[|\psi_x\rangle\langle\psi_x|] = \frac{\mathbb{I}}{D}$ (since the Hadamard basis states form an orthonormal basis):

$$\begin{aligned}\mathbb{E}_x\Big[\mathbb{1}[f(x)\in\mathcal{B}]\cdot\mathrm{Tr}\Big(M_{f(x)}|\psi_x\rangle\langle\psi_x|\Big)\Big] &\le \mathbb{E}_x\left[\sum_{i\in\mathcal{B}}\mathrm{Tr}(M_i|\psi_x\rangle\langle\psi_x|)\right] \\ &= \sum_{i\in\mathcal{B}}\mathrm{Tr}\left(M_i\cdot\frac{\mathbb{I}}{D}\right) = \frac{1}{D}\sum_{i\in\mathcal{B}}\mathrm{Tr}(M_i).\end{aligned} \tag{22}$$

To bound $\sum_{i\in\mathcal{B}}\mathrm{Tr}(M_i)$, we use the global trace constraint $\sum_i \mathrm{Tr}(M_i) = D$ (for a complete POVM on a $D$-dimensional space) and the lower bound $\mathrm{Tr}(M_i) \ge M_i[i,i] \ge 1-\sqrt{\epsilon}$ for $i\in\mathcal{G}$:

$$\begin{aligned}\sum_{i\in\mathcal{B}}\mathrm{Tr}(M_i) &= D - \sum_{i\in\mathcal{G}}\mathrm{Tr}(M_i) \le D - |\mathcal{G}|\,(1-\sqrt{\epsilon}) \\ &= D-(D-|\mathcal{B}|)(1-\sqrt{\epsilon}) = |\mathcal{B}| + (D-|\mathcal{B}|)\sqrt{\epsilon} \le |\mathcal{B}| + D\sqrt{\epsilon}.\end{aligned} \tag{23}$$

Since $|\mathcal{B}| \le D\sqrt{\epsilon}$, we have $\sum_{i\in\mathcal{B}}\mathrm{Tr}(M_i) \le 2D\sqrt{\epsilon}$, and thus:

$$\mathbb{E}_x\Big[\mathbb{1}[f(x)\in\mathcal{B}]\cdot\mathrm{Tr}\Big(M_{f(x)}|\psi_x\rangle\langle\psi_x|\Big)\Big] \le \frac{1}{D}\cdot 2D\sqrt{\epsilon} = 2\sqrt{\epsilon}. \tag{24}$$

**Combining:** The total guessing probability is bounded by:

$$\mathbb{E}_x\Big[\mathrm{Tr}\Big(M_{f(x)}|\psi_x\rangle\langle\psi_x|\Big)\Big] \le \frac{1}{D} + 2\epsilon^{\frac{1}{4}} + 2\sqrt{\epsilon} \le \frac{1}{D} + 4\epsilon^{\frac{1}{4}} \tag{25}$$

for $\epsilon \le 1$ (using $\sqrt{\epsilon} \le \epsilon^{\frac{1}{4}}$). Since this holds for all functions $f$, the maximum is also bounded by $\frac{1}{D} + O\big(\epsilon^{\frac{1}{4}}\big)$. ■

# 4 Scheme and Classical-Query Security

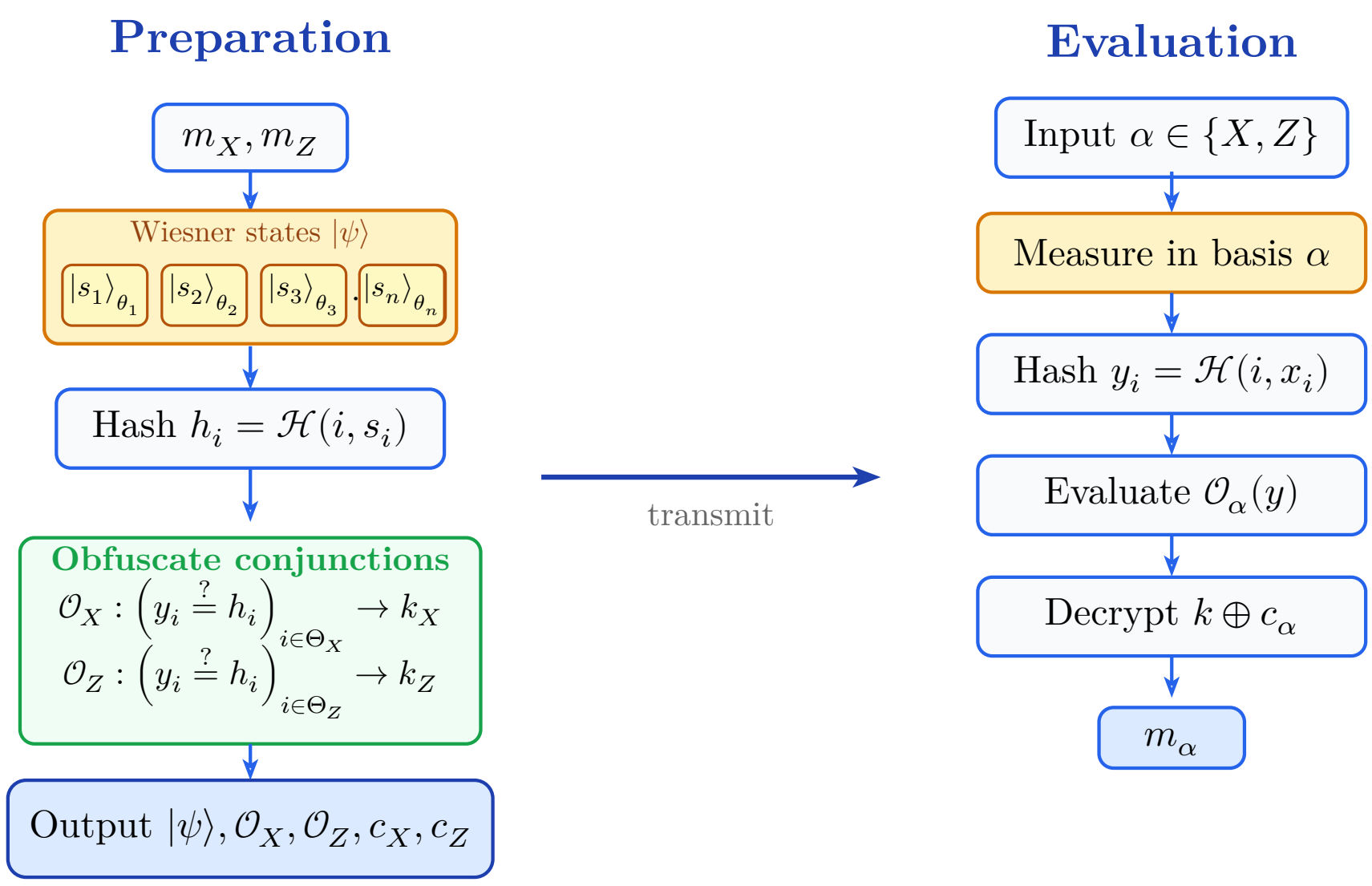

Figure 2: Construction overview. Quantum operations shown in yellow.

We now construct our one-time memories and establish its security against quantum adversaries restricted to classical random oracle queries. The lifting to full quantum depth-bounded security is deferred to Section 5.

The construction employs two primitives: tensor products of independent Wiesner states (with no entanglement between positions) and distributional VBB obfuscation for conjunctions [3] (recalled in Section 2).

## 4.1 The Construction

We present the one-time program scheme. The construction operates in the random oracle model with a single random oracle $\mathcal{H} : [n] \times \{0,1\}^m \to \{0,1\}^{m'}$ for $m' = \Theta(\lambda)$ (to ensure collision resistance). We use $H$ to denote the single-qubit Hadamard gate throughout.

**Program 1:** Conjunction Program $C_\alpha$

**Input:** $y = (y_1, ..., y_n) \in \left(\{0,1\}^{m'}\right)^n$

**Hard-coded:** Positions $\Theta_\alpha \subseteq [n]$, hashes $\{h_i\}_{i \in \Theta_\alpha}$, key $k_\alpha \in \{0,1\}^\lambda$

---

**if** $\forall i \in \Theta_\alpha : h_i = y_i$ **then return** $k_\alpha$

**else return** $\perp$

**Scheme 2:** One-Time Program Scheme Π

**Parameters:** Security parameter $\lambda$, messages $m_X, m_Z \in \{0,1\}^\lambda$

**Preparation** OTM.Prep$(1^\lambda, m_X, m_Z)$:

Set $n = \tilde{\Theta}(\lambda)$, $m = \tilde{\Theta}(\lambda)$

For $i \in [n]$: sample $\theta_i \leftarrow \{X, Z\}$ and $s_i \leftarrow \{0,1\}^m$ uniformly

Let $\Theta_X = \{i : \theta_i = X\}$ and $\Theta_Z = \{i : \theta_i = Z\}$

Prepare state: $|\psi\rangle = \bigotimes_{i=1}^n \left[(H^{\otimes m})|s_i\rangle \text{ if } \theta_i = X, \text{ else } |s_i\rangle\right]$

Compute hashes: $h_i = \mathcal{H}(i, s_i)$ for all $i \in [n]$

Sample keys: $k_X, k_Z \leftarrow \{0,1\}^\lambda$ uniformly

Compute ciphertexts: $c_\alpha = k_\alpha \oplus m_\alpha$ for $\alpha \in \{X, Z\}$

Generate: $\mathcal{O}_\alpha = \text{ObfConj}(1^\lambda, C_\alpha)$ with hashes $\{h_i\}_{i \in \Theta_\alpha}$ and key $k_\alpha$

**Output:** $(|\psi\rangle, \mathcal{O}_X, \mathcal{O}_Z, c_X, c_Z)$

**Evaluation** OTM.Eval$(|\psi\rangle, \mathcal{O}_X, \mathcal{O}_Z, c_X, c_Z, \alpha)$:

Measure $|\psi\rangle$ in basis $\alpha$, obtaining $(x_1, ..., x_n)$

Compute $y_i = \mathcal{H}(i, x_i)$ for all $i \in [n]$; let $y = (y_1, ..., y_n)$

Compute $k = \mathcal{O}_\alpha(y)$; if $k = \bot$ output $\bot$

**Output:** $k \oplus c_\alpha$

## 4.2 Correctness and Security

We now establish the correctness and security properties of the scheme.

**Lemma 4.1** *(Correctness)*: Let Π denote the protocol specified in Scheme 2. For any basis choice $\alpha \in \{X, Z\}$, an honest evaluator recovers $m_\alpha$ with probability 1.

*Proof*: An honest evaluator measuring in basis $\alpha$ obtains the correct value $s_i$ for all positions $i \in \Theta_\alpha$, since the Wiesner states at these positions are encoded in basis $\alpha$. Computing $y_i = \mathcal{H}(i, s_i)$ yields the correct hashes $h_i$ for $i \in \Theta_\alpha$, so the conjunction $\mathcal{O}_\alpha$ accepts and returns $k_\alpha$. The evaluator then computes $k_\alpha \oplus c_\alpha = k_\alpha \oplus (k_\alpha \oplus m_\alpha) = m_\alpha$.■

The security analysis relies on three properties that together prevent an adversary from recovering both messages:

1. POVM bound (Section 3): Any measurement achieving high success probability on one basis yields post-measurement states that are close to maximally mixed in the conjugate basis.
2. Random oracle extraction: The random oracle $\mathcal{H}$ forces the adversary to *classically* query $\mathcal{H}(i, s_i)$ for each position $i$ where they wish to provide a valid input to the conjunction, enabling extraction of measurement outcomes.

3. Conjunction hiding: The distributional VBB security of ObfConj ensures that the positions $\Theta_\alpha$ and the key $k_\alpha$ remain hidden from the adversary until the conjunction accepts, preventing basis-selective measurement strategies.

### 4.2.1 Classical-Query Security

We now state and prove the main security theorem for the classical-query setting.

**Theorem 4.1** (Classical-Query Security): Assume the LPN assumption holds. The scheme $\Pi$ from Scheme 2 is simulation-secure in the random oracle model against quantum adversaries that are restricted to classical queries to the random oracle $\mathcal{H}$.

*Proof*: We construct a simulator $\mathcal{S}$ and establish indistinguishability via a sequence of hybrid games.

**Simulator Construction.** The simulator $\mathrm{Sim}^{g^{m_X, m_Z}}$ operates as follows:

1. Sample the Wiesner state $|\psi\rangle$, obfuscations $\mathcal{O}_X, \mathcal{O}_Z$, and keys $k_X, k_Z$ as in the real protocol. Set $c_X, c_Z \leftarrow \{0,1\}^\lambda$ uniformly.
2. Run $\mathcal{A}$ on $(|\psi\rangle, \mathcal{O}_X, \mathcal{O}_Z, c_X, c_Z)$, programming the random oracle lazily. Since the simulator knows the secrets $s_i$, it monitors classical queries to $\mathcal{H}$ and detects which positions the adversary queries correctly, thereby extracting the adversary's basis choice $\alpha$.
3. Query $m_\alpha \leftarrow g^{m_X, m_Z}(\alpha)$ and set $c_\alpha := k_\alpha \oplus m_\alpha$ (the simulator controls the VBB-simulated obfuscation and can patch the conjunction output so that $\mathcal{A}$ obtains $m_\alpha$ upon decryption). Sample $c_{\overline{\alpha}} \leftarrow \{0,1\}^\lambda$ uniformly (already set in step 1).
4. Output $\mathcal{A}$'s view.

We now present a sequence of hybrids. Let Win denote the event that the adversary can distinguish the real execution from the simulated one.

**Hybrid $H_0$ (Real Execution):** The honest protocol execution with $\mathrm{OTM.Prep}(1^\lambda, m_X, m_Z)$.

**Hybrid $H_1$ (Simulated Obfuscation):** Identical to $H_0$, except $\mathcal{O}_X, \mathcal{O}_Z$ are replaced with simulated conjunction obfuscations $\tilde{\mathcal{O}}_X, \tilde{\mathcal{O}}_Z$ from the VBB simulator $\mathcal{S}_{\mathrm{VBB}}$ with oracle access to the respective $C_\alpha$ functions.

We note that this follows by the distributional VBB security under LPN [3][6]. Thus:

$$|\Pr[\mathrm{Win} \mid H_0] - \Pr[\mathrm{Win} \mid H_1]| \leq \mathrm{negl}(\lambda). \tag{26}$$

**Hybrid $H_2$ (Simulator):** Identical to $H_1$, except $c_{\overline{\alpha}}$ is sampled uniformly at random (independent of $m_{\overline{\alpha}}$), while $c_\alpha = k_\alpha \oplus m_\alpha$ remains correct.

[6]Recall that we conjecture that distributional VBB security extends to quantum auxiliary inputs; see Section 2 for the formal statement.

The transition $H_1 \to H_2$ relies on the one-time property: the adversary can evaluate at most one conjunction. To evaluate $\mathcal{O}_{\overline{\alpha}}$, the adversary must query $\mathcal{H}(i, s_i)$ for all $i \in \Theta_{\overline{\alpha}}$, which requires guessing all conjugate-basis secrets.

Crucially, each hash query $\mathcal{H}(i, x_i)$ requires a *classical* input $x_i$, irrevocably committing the adversary to outcome $x_i$ at position $i$. Since the $n$ Wiesner states are prepared as an independent tensor product $\bigotimes_i |s_i\rangle_{\theta_i}$, and the hash is keyed per position, the adversary's success probability at each position is independent conditioned on its committed outcomes. By Theorem 3.1 applied independently at each position (and its symmetric analogue, obtained by conjugating with $H^{\otimes m}$, for the case $\alpha = X$), if $\mathcal{A}$ successfully queries all positions in $\Theta_\alpha$ (measuring in basis $\alpha$ with per-position POVM error $\epsilon_i$), the probability of also guessing each conjugate-basis secret is at most $2^{-m} + O\left(\epsilon_i^{\frac{1}{4}}\right)$, where $\epsilon_i$ must be negligible for the conjunction to accept. Taking the product over $|\Theta_{\overline{\alpha}}| = \tilde{\Theta}(\lambda)$ independent positions with $m = \tilde{\Theta}(\lambda)$:

$$\Pr[\mathcal{A} \text{ evaluates both conjunctions}] \leq 2^{-\Omega(\lambda^2)} = \operatorname{negl}(\lambda). \tag{27}$$

Since $\mathcal{A}$ cannot recover $k_{\overline{\alpha}}$ except with negligible probability, the value $c_{\overline{\alpha}} = k_{\overline{\alpha}} \oplus m_{\overline{\alpha}}$ is indistinguishable from uniform. Thus $H_2 \approx H_1 \approx H_0$, and $H_2$ is exactly the simulator's output. ■

# 5 Lifting to Quantum Depth Security

The scheme in Section 4 is secure against adversaries that can only make *classical* queries to the random oracle $\mathcal{H}$ (Theorem 4.1). We now lift this to security against adversaries with bounded *quantum* depth using the lifting framework of Arora et al. [1].

Recall from Definition 2.6 that a depth-$d$ bounded adversary is modeled by the complexity class $\mathsf{BPP}^{\mathsf{QNC}_d^{\mathsf{BPP}}}$: the adversary performs polynomial-time classical computation interleaved with depth-$d$ quantum circuits which can in turn invoke interleaved classical computation.

We informally adapt the framework of Arora et al. [1] (specifically, Lemma 6) to argue that depth-bounded security follows from classical-query security. At a high level, their result establishes that for any $\mathsf{BPP}^{\mathsf{QNC}_d^{\mathsf{BPP}}}$ adversary $\mathcal{A}$ with $d = \operatorname{poly}(\lambda)$ interacting with a random oracle, there exists a modified oracle $\widetilde{\mathcal{H}}$ such that: (1) $\mathcal{A}$ cannot distinguish $\widetilde{\mathcal{H}}$ from a true random oracle, and (2) for a problem $\mathcal{P}$ which is classical-query sound relative to $\mathcal{H}$, one can replace $\mathcal{H}$ with $\widetilde{\mathcal{H}}$ to obtain a problem $\mathcal{P}'$ that is sound against $\mathsf{BPP}^{\mathsf{QNC}_d^{\mathsf{BPP}}}$ adversaries.

We note that Arora et al. formalize a "problem" $\mathcal{P}$ as a relation between an adversary's view and a secret value, and define classical-query soundness as the inability of classical-query adversaries to produce valid views for $\mathcal{P}$ with non-negligible probability. The lifting theorem then constructs a modified oracle $\widetilde{\mathcal{H}}$ such that any depth-$d$ quantum adversary's interaction with $\widetilde{\mathcal{H}}$ effectively collapses to a classical transcript, allowing the classical-query soundness of $\mathcal{P}$ to carry over to depth-bounded soundness of $\mathcal{P}'$.

The theorem leverages the observation that depth-bounded adversaries cannot maintain coherent superpositions across many oracle queries without intermediate measurements collapsing the state. We recommend Ref. [1] for details.

**Conjecture 5.1** *(Simulation-Security against* $\mathsf{BPP}^{\mathsf{QNC}_d^{\mathsf{BPP}}}$ *Adversaries)*: The scheme from Scheme 2 is simulation-secure against $\mathsf{BPP}^{\mathsf{QNC}_d^{\mathsf{BPP}}}$ adversaries for any $d = \operatorname{poly}(\lambda)$.

Assuming the lifting framework of Arora et al. applies to our setting, we can replace the problem of an adversary recovering inputs $\{h_i\}_{i\in\Theta_\beta}$ with $\beta = X$ if $\alpha = Z$ and vice versa in the classical-query setting with the same problem in the depth-bounded quantum-query setting. Thus, the simulator in the proof of Theorem 4.1 can replace $c_\beta$ with a random value as $k_\beta$ remains hidden from depth-bounded quantum adversaries. A full formal proof would require taking the informal lifting lemma in Ref. [1] and adapting it to fit our setting; we leave this to future work.

# 6 Conclusion and Future Directions

In this work, we present a one-time memory (OTM) construction secure against quantum adversaries with polynomially bounded adaptive depth. Our approach combines three ingredients: (1) single-qubit Wiesner states, (2) conjunction obfuscation, and (3) the Arora et al. depth-bounded lifting framework [1]. The key technical contribution is proving classical-query security, which we conjecture extends to the full depth-bounded setting via the Arora et al. lifting framework. The resulting scheme is conceptually simple and points toward practical implementation using near-term quantum hardware.

Several directions remain open:

- **Tighter POVM bounds**: Our current bound of $O\left(\epsilon^{\frac{1}{4}}\right)$ may be improvable. Can we achieve $O\left(\epsilon^{\frac{1}{2}}\right)$ or better with a more refined analysis?
- **Noise tolerance**: Analyzing the scheme's robustness to realistic quantum noise (depolarizing, dephasing) would strengthen the practical applicability.
- **Formalizing memory without fault tolerance:** Ideally, we would like to have a notion of encoding quantum states into memory which are incapable of universal fault-tolerant computation, but still allow for some limited quantum processing. Formalizing this notion and proving security in this model would capture realistic adversaries and the full usefulness of one-time programs.
- **Proving the lifting theorem in our setting:** While we provided an informal argument for applying the lifting theorem of Arora et al. [1] to our setting, a formal proof would be needed prior to real deployment of our scheme. Though conceptually simple, we note the technical details of the proof in Ref. [1] are quite involved. We thus leave this to future work.

# Acknowledgments

The author is grateful to the helpful discussions and feedback from Fabrizio Romano Genovese, Stefano Gogioso, and Matthew Coudron. The author also acknowledges funding and support from NeverLocal Ltd, Neon Tetra LLC, and from the NSF Graduate Research Fellowship Program.

**AI usage:** the author would like to acknowledge the use of language models, Gemini and Claude, for assistance with editing and typesetting.